\documentclass[preprint, showpacs, preprintnumbers, amsmath, amssymb]{revtex4}
\usepackage{graphicx}
\usepackage{dcolumn}
\usepackage{bm}

\begin{document}

\preprint{...}

\title{Novel Topological
Invariant in the $U(1)$ Gauge Field Theory}
\author{Yi-Shi Duan}
\author{Xin-Hui Zhang}
\email{zhangxingh03@st.lzu.edu.cn}
\author{Li Zhao}
\affiliation{
Institute of Theoretical Physics, Lanzhou University \\
Lanzhou 730000, People's Republic of China}

\begin{abstract}
Based on the decomposition of $U(1)$ gauge potential theory and the
$\phi$-mapping topological current theory, the three-dimensional knot
invariant and a four-dimensional new topological invariant are discussed in
the $U(1)$ gauge field.

\end{abstract}
\maketitle

\section{introduction}

Knotlike configurations appear in a variety of physical, chemical and
biological scenarios, including the structure of elementary particles
\cite{67,10,9}, the early universe cosmology \cite{liuxin64,51,62}, the
Bose-Einstein condensation \cite{liuxin3,12}, the polymer folding
\cite{liuxin86} and the DNA replication, transcription and recombination
\cite{liuxin48}. Generally, knotlike configurations are researched by using
the Jones polynomial and path integral methods in physics
\cite{jones,king3,king4}. Meanwhile, it is known that for a knot family
there are very important characteristic numbers to describe its topology
\cite{qq}. So in the research of the knotlike configurations in physics,
one should also pay much attention to the geometric and topological
characteristics.

As is well known, magnetic helicity has attracted a great deal of attention
since it provides a lower bound for the magnetic energy
\cite{link3,link4,link6} and is a very robust invariant. Magnetic helicity
can be considered as a special case of the cross helicity, which is defined
for two distinct magnetic fields $B$ and $B'$
\begin{equation}\label{H}
H(B,B')=\int_{M^3}A\wedge{dA'}.
\end{equation}
So the question arises whether there exists a generalization to
electromagnetic field, especially to the physically interesting
four-dimensional case. Principally, the electromagnetic field can be
investigated in the frame of a $U(1)$ gauge field theory. In this paper, we
intend to study the geometric and topological characteristic of knotlike
vortex lines in the electromagnetic field by using the decomposition of
$U(1)$ gauge potential theory and the $\phi$-mapping theory. The gauge
potential decomposition theory \cite{liuxin46} establishes a direct
relationship between differential geometry and topology, and the
$\phi$-mapping topological current theory provides an important method in
investigating the inner topological structure of field configurations
\cite{26,liuxin13,liuxin15}. The purpose of the present paper is twofold.
First we show that there exist vortex lines in the $U(1)$ gauge field and
obtain a knot invariant for knotlike vortex lines. The second purpose is to
extend to four-dimensional case and give a new topological invariant.

This paper is arranged as follows. In Sec. \ref{section2}, using the $U(1)$
gauge potential decomposition theory and the $\phi$-mapping topological
current theory, we discuss the inner topological structure of the knotlike
vortex lines in the $U(1)$ gauge field and get a knot invariant. In Sec.
\ref{section3}, a new topological invariant is obtained in a
four-dimensional case. The conclusion of this paper is given in Sec.
\ref{section4}.

\section{knot invariant in the $U(1)$ gauge field}\label{section2}

We know that the complex scalar field $\phi$ can be regarded as the complex
representation of a two-dimensional vector field $\vec{\phi}=(\phi^{1},
\phi^{2})$ over the base space, i.e. $\phi=\phi^{1}+i\phi^{2}$, where
$\phi^{a}\;(a=1,2)$ are real functions. The covariant derivative of $\phi$
is defined as
\begin{equation}
D_{\mu}\phi=\partial_{\mu}\phi-iA_{\mu}\phi,\;\;\;(\mu=1, 2, 3),
\end{equation}
where $A_{\mu}$ is the $U(1)$ gauge potential. The $U(1)$ gauge field
tensor is given by
$F_{\mu\nu}=\partial_{\mu}A_{\nu}-\partial_{\nu}A_{\mu}$. In the subsection
we will show that in the $U(1)$ gauge field theory there exist the vortex
line structures. Defining the unit vector $n^a$ as
\begin{equation}
n^{a}=\frac{\phi^{a}}{\|\phi\|},\;\;\;\|\phi\|^{2}=\phi\phi^{*},
\end{equation} one can prove that \cite{liuxin13},
$A_{\mu}$ can be decomposed in terms of $n^{a}$:
$A_{\mu}=\epsilon^{ab}n^{a}\partial_{\mu}n^{b}-\partial_{\mu}\theta$, where
$\theta$ is a phase factor. Since the term $(\partial_{\mu}\theta)$  dose
not contribute to the gauge field tensor, $A_{\mu}$ and $F_{\mu\nu}$ can be
rewritten as
\begin{equation}\label{potential1}
A_{\mu}=\epsilon^{ab}n^{a}\partial_{\mu}n^{b},\;\;\;
F_{\mu\nu}=2\epsilon^{ab}\partial_{\mu}n^{a}\partial_{\nu}n^{b}.
\end{equation}
In the following, we will use these above decomposition expressions Eq.
(\ref{potential1}) to research the topological feature in the $U(1)$ gauge
field theory. As an analog of Eq. (\ref{H}), define a action $I$ in terms
of two distinct electromagnetic field tensors $F^i$ and $F^j$ by
\begin{equation}\label{chernsimon}
I(F^i,F^j)=\frac{1}{4\pi}\int_{M}A^i\wedge{F^j}
=\frac{1}{8\pi}\int_{M}\epsilon^{\mu\nu\lambda}A^i_{\mu}F^j_{\nu\lambda}d^{3}x.
\end{equation}
Substituting $F^i=F^j$ in Eq. (\ref{chernsimon}) returns the expression for
Chern-Simon action \cite{9}. To simplify the action, we introduce a
topological current
\begin{equation}\label{Jmunu1}
J^{\mu}=\frac{1}{8\pi}\epsilon^{\mu\nu\lambda}F^j_{\nu\lambda}.
\end{equation}
Then Eq. (\ref{chernsimon}) can be expressed as
\begin{equation}\label{chernsimon1}
I(F^i,F^j)=\int_{M}A^i_{\mu}J^{\mu}d^{3}x.
\end{equation}
In order to explore the inner topological structure of the action $I$, we
should first investigate that of $J^{\mu}$. Then taking account of Eq.
(\ref{potential1}), using
$\partial_{\mu}n^{a}=\partial_{\mu}\phi^{a}/\|\phi\|+\phi^{a}\partial_{\mu}(1/\|\phi\|)$
and the Green's function relation in $\phi$ space \cite{liuxin15}, one can
prove that
\begin{equation}\label{Jmunu2}
J^{\mu}=\delta^{2}(\vec{\phi})D^{\mu}(\frac{\phi}{x}),
\end{equation}
where
\begin{equation}\label{Jacobian1}
D^{\mu}(\frac{\phi}{x})=\frac{1}{2}\epsilon^{\mu\lambda\rho}
\epsilon^{ab}\partial_{\lambda}\phi^{a}\partial_{\rho}\phi^{b}
\end{equation}
is the Jacobian vector. The expression Eq. (\ref{Jmunu2}) provides an
important conclusion:
\begin{equation}
J^{\mu}\left\{
\begin{array}{c}
=0 \\
\neq0
\end{array}
\right.
\begin{array}{c}
\textrm{if and only if}\;\;\vec{\phi}\neq0. \\
\textrm{if and only if}\;\;\vec{\phi}=0.
\end{array}
\end{equation}
So it is necessary to study the zero points of the $\vec\phi$ field to
determine the nonzero solutions of $J^{\mu}$. The implicit function theory
\cite{implicit} shows  that under the regular condition
$D^{\mu}(\phi/x)\neq0$, the general solutions of
\begin{equation}
\phi^{1}(x^{1},x^{2},x^3)=0,\;\;\;\phi^{2}(x^{1},x^{2},x^3)=0
\end{equation}
can be expressed as
\begin{equation}
x^{1}=x^{1}_{k}(s),\;\;\; x^{2}=x^{2}_{k}(s),\;\;\;x^{3}=x^{3}_{k}(s)
\end{equation}
which represent $N$ isolated singular strings $L_{k}$ with string parameter
$s\;(k=1,2,...,N)$. These singular string solutions are just the vortex
lines  in the $U(1)$ gauge field.

In the $\delta$-function theory \cite{delta}, one can prove that in
three-dimensional space
\begin{equation}\label{delta}
\delta^{2}(\vec{\phi})=\sum^{N}_{k=1}\beta_{k}\int_{L_{k}}
\frac{\delta^{3}(\vec{x}-\vec{x}_{k}(s))}{|D(\frac{\phi}{u})|_{\Sigma_{k}}}ds,
\end{equation}
where
\begin{equation}\label{Jacobian2}
D(\phi/u)_{\Sigma_{k}}=\frac{1}{2}\epsilon^{\mu\nu}\epsilon_{mn}
(\partial\phi^{m}/\partial{u^{\mu}})(\partial\phi^{n}/\partial{u^{\nu}}),
\end{equation}
and $\Sigma_{k}$ is the $k$th planar element transverse to $L_{k}$ with
local coordinates $(u^{1}, u^{2})$. The positive integer $\beta_{k}$ is the
Hopf index of the $\phi$-mapping, which means that when $\vec{x}$ covers
the neighborhood of the zero point $\vec{x}_{k}(s)$ once, the vector field
$\vec\phi$ covers the corresponding region in $\phi$ space $\beta_{k}$
times. Meanwhile taking notice of Eqs. ({\ref{Jacobian1}}) and
({\ref{Jacobian2}}), the direction vector of $L_{k}$ is given
\cite{liuxin15} by
\begin{equation}\label{volicity}
\left.\frac{dx^{\mu}}{ds}\right|_{\vec{x}_{k}}
=\left.\frac{D^{\mu}(\phi/x)}{D(\phi/u)}\right|_{\vec{x}_{k}}.
\end{equation}
Then from Eqs. (\ref{delta}) and (\ref{volicity}), we obtain the inner
structure of $J^{\mu}$
\begin{equation}\label{ji2}
J^{\mu}=\delta^{2}(\vec{\phi})D^{\mu}(\frac{\phi}{x})
=\sum^{N}_{k=1}W_{k}\int_{L_{k}}\frac{dx^{\mu}}{ds}\delta^{3}(\vec{x}-\vec{x}_{k}(s))ds,
\end{equation}
in which $W_{k}=\beta_{k}\eta_{k}$ is the winding number of $\vec{\phi}$
around $L_{k}$, with $\eta_{k}=\textrm{sgn}D(\phi/u)_{\vec{x}_{k}}=\pm1$
being the Brouwer degree of $\phi$-mapping. It can be seen that when these
$U(1)$ vortex lines are $N$ closed curves, i.e., a family of $N$ knots
$\gamma_{k} (k=1,...,N)$, and taking account of Eqs. (\ref{chernsimon1})
and (\ref{ji2}), the action $I(F^i,F^j)$ is reexpressed as
\begin{equation}\label{chernsimon2}
I(F^i,F^j)=\int_{M}A^i_{\mu}J^{\mu}dx^{3}
=\sum^{N}_{k=1}W_{k}\oint_{\gamma_{k}}A^i_{\mu}dx^{\mu}.
\end{equation}
This is a very important expression. Consider the $U(1)$ gauge
transformation of $A_{\mu}$:
\begin{equation}\label{$U(1)$}
{A^i}'_{\mu}=A^i_{\mu}+\partial_{\mu}\theta,
\end{equation} where $\theta$
is a phase factor denoting the $U(1)$ transformation. It can be shown that
the $(\partial_{\mu}\theta)$ term in Eq. (\ref{$U(1)$}) contributes nothing
to the integral $I$, that is, the expression Eq. (\ref{chernsimon2}) is
invariant under the $U(1)$ gauge transformation. Meanwhile we know that $I$
is independent of the metric $g_{\mu\nu}$. Therefore one can conclude that
the action $I(F^i,F^j)$ in Eq. (\ref{chernsimon}) is a topological
invariant for the knotlike vortex lines in the $U(1)$ gauge field theory.

\section{the new four-dimensional topological invariant}\label{section3}

The action $I (F^i,F^j)$ is of three-dimensional nature due to the integral
over $M^{3}$, so the question arises whether there exists a generalization
to higher dimensions, especially to the physically interesting
four-dimensional case \cite{2004prl}. In this section, we construct a new
action in four dimensions, which is based on the so-called
Novikov invariant \cite{Novikov} .\\
\indent In the following the field tensor
$F_{\mu\nu}=\partial_{\mu}A_{\nu}-\partial_{\nu}A_{\mu}$ is also
interpreted as the $U(1)$ gauge field over a simply connected domain
$M^{4}$, in which $A_{\mu}$ is the $U(1)$ gauge potential. The new action
is defined in terms of three distinct electromagnetic field tensor $F^i$,
$F^j$ and $F^k$ \cite{2004prl}
\begin{equation}\label{linking}
N(F^i,F^j,F^k)=\frac{1}{4\pi}\int_{M^{4}}{A^i}\wedge{A^j}\wedge{F^k}
=\frac{1}{8\pi}\int_{M^{4}}
\epsilon^{\mu\nu\lambda\rho}A^i_{\mu}A^j_{\nu}F^k_{\lambda\rho}d^{4}x,
\end{equation}
where $\mu,\nu,\lambda,\rho=1,2,3,4$. And then we will research the inner
topological structure of the new action. Introduce a two-dimensional
topological tensor current, which is denoted as \cite{2current}
\begin{equation}\label{Jmunu1}
J^{\mu\nu}=\frac{1}{8\pi}\epsilon^{\mu\nu\lambda\rho}F^k_{\lambda\rho}.
\end{equation}
Substituting Eq. (\ref{Jmunu1}) into Eq. (\ref{linking}), the new action
can be simplified as
\begin{equation}\label{linking1}
N(F^i,F^j,F^k)=\int_{M^{4}}A^i_{\mu}A^j_{\nu}J^{\mu\nu}d^{4}x.
\end{equation}
To explore the inner topological structure of the action $N(F^i,F^j,F^k)$,
we research that of $J^{\mu\nu}$. Using the $\phi$-mapping topological
theory \cite{liuxin15}, we can obtain
\begin{equation}\label{JJ}
J^{\mu\nu}=\delta^{2}(\vec{\phi})D^{\mu\nu}(\frac{\phi}{x}),
\end{equation}
where
\begin{equation}\label{jacobian3}
D^{\mu\nu}(\frac{\phi}{x})=\frac{1}{2}\epsilon^{\mu\nu\lambda\rho}\epsilon^{ab}
\partial_{\lambda}\phi^{a}\partial_{\rho}\phi^{b}.
\end{equation}
Considering the Eq. ({\ref{JJ}}), we can come to an important conclusion
\begin{equation}
J^{\mu\nu}\left\{
\begin{array}{c}
=0 \\
\neq0
\end{array}
\right.
\begin{array}{c}
\textrm{if and only if}\;\;\vec{\phi}\neq0. \\
\textrm{if and only if}\;\;\;\vec{\phi}=0.
\end{array}
\end{equation}
Now we should first study the zero points of $\vec\phi$ to determine the
nonzero solutions of $J^{\mu\nu}$. The implicit function theory
\cite{implicit} shows that under the regular condition
\begin{equation}
D^{\mu\nu}(\phi/x)\neq0,
\end{equation}
the general solutions of
\begin{equation}
\phi^{1}(x^{1},x^{2},x^{3},x^{4})=0,\;\;\;
\phi^{2}(x^{1},x^{2},x^{3},x^{4})=0
\end{equation}
can be expressed as
\begin{equation}
x^{1}=x^{1}_{l}(\sigma),\;\; x^{2}=x^{2}_{l}(\sigma),\;\;
x^{3}=x^{3}_{l}(\sigma),\;\; x^{4}=x^{4}_{l}(\sigma),
\end{equation}
which represent  $N$ isolated singular sheets with surface parameter
$\sigma$. Without loss of generality, we consider these singular sheet
solutions bordering by two vortex lines with string parameter $s\;
(l=1,2,...,N)$. Generalizing Eq. (\ref{delta}) to four-dimensional case
with two vortex lines $L_{k}$ and $L_{\gamma}$, $\delta({\vec\phi})$ can be
expressed as
\begin{equation}\label{delta2}
\delta^{2}(\vec{\phi})=\sum^{N}_{k=1}\sum^{N}_{l=1}\beta_{k}\beta_{l}\int_{L_{k}}
\int_{L_{l}}\frac{\delta^{2}(\vec{x}-\vec{x}_{k}(s))\delta^{2}(\vec{x}-\vec{x}_{l}(s))}
{|D(\frac{\phi}{u})|_{\Sigma_{k}}|D(\frac{\phi}{v})|_{\Sigma_{l}}}d^{2}s,
\end{equation}
where $\Sigma_{k}$ is the $k$th planar element transverse to $L_{k}$ with
local coordinates $(u^{1},u^{2})$ and $\Sigma_{l}$ is transverse to $L_{l}$
with $(v^{1},v^{2})$. Taking notice of Eqs. ({\ref{Jacobian2}}) and
({\ref{jacobian3}}), we can obtain
\begin{equation}\label{volicity2}
\left.\frac{dx^{\mu}\wedge{dx^{\nu}}}{d^{2}s}\right|_{\vec{x}_{k},\vec{x}_{l}}=\frac{D^{\mu\nu}
(\phi/x)}{D(\phi/u)|_{\Sigma_{k}}D(\phi/v)|_{\Sigma_{l}}}.
\end{equation}
Then from Eqs. (\ref{delta2}) and  (\ref{volicity2}), we get the inner
structure of $J^{\mu\nu}$
\begin{eqnarray}
J^{\mu\nu}&=&\delta^{2}(\vec{\phi})D^{\mu\nu}(\frac{\phi}{x})\\\nonumber
&=&\sum^{N}_{k=1}\sum^{N}_{k=1}W_{k}W_{l}\int_{L_{k}}\int_{L_{l}}
\frac{dx^{\mu}\wedge{dx^{\nu}}}{d^{2}s}\\\nonumber
&&\delta^{2}(\vec{x}-\vec{x}_{k}(s))\delta^{2}(\vec{x}-\vec{x}_{l}(s))d^{2}s,
\end{eqnarray}
in which $W_{k}=\beta_{k}\eta_{k}$ is the winding number of $\vec\phi$
around $L_k$ with
$\eta_{k}=\textrm{sgn}D(\frac{\phi}{u})_{\Sigma_{k}}=\pm1$ being the Brower
degree of the $\phi$-mapping, while $W_{l}=\beta_{l}\eta_{l}$ is the
winding number corresponding to $L_l$. When these vortex lines are closed
curves, i.e. a family of knots
$\gamma_{k},\;\gamma_{l}\;\;(k,l=1\cdot\cdot\cdot{N})$, the new action Eq.
(\ref{linking1}) can be expressed as
\begin{equation}\label{linking3}
N(F^i,F^j,F^k)=\sum^{N}_{k=1}\sum^N_{l=1}W_{k}W_{l}
\oint_{\gamma_{k}}A^i_{\mu}dx^{\mu}\wedge\oint_{\gamma_{l}}A^j_{\nu}dx^{\nu}.
\end{equation}
Considering the $U(1)$ transformation of $A$
\begin{equation}
{A^i}'_{\mu}=A^i_{\mu}+\partial_{\mu}\alpha,\;\;\;
{A^j}'_{\nu}=A^j_{\nu}+\partial_{\nu}\theta,
\end{equation}
one can seen that the $\partial_{\mu}\alpha$ and $\partial_{\nu}\theta$
contribute nothing to the action $N(F^i,F^j,F^k)$. So the expression Eq.
(\ref{linking3}) is invariant under the gauge transformation. Meanwhile we
know that $N(F^i,F^j,F^k)$ is independent of the metric. Therefore one can
conclude that $N(F^i,F^j,F^k)$ is a topological invariant. Since the
invariant $N(F^i,F^j,F^k)$ is concerned with two families of knotlike
vortex lines $\gamma_{k}$ and $\gamma_{l}\;(k,\;l=1,\cdot\cdot\cdot{N})$,
we conjecture that it is invariant for the linkage of vortex lines. As
follows, we will research the relationship between the new invariant and
the linking numbers of the knots family. We should first express $A_{\mu}$
in terms of the vector field which carries the geometric information of the
linkage, namely, we need to decompose $A^i_{\mu}$ and $A^j_\mu$ in terms of
another two-dimensional unit vector $\vec{e}$ which is different from the
two-dimensional vector $\vec{n}$ in Sec. II. Define the Gauss mapping
$\vec{m}$
\begin{equation}\label{m}
\vec{m}(\vec{y},\vec{x})=\frac{\vec{y}-\vec{x}}{||\vec{y}-\vec{x}||},
\end{equation}
where $\vec{x}$ and $\vec{y}$ are two points, respectively, on the knots
$\gamma_{k}$ and $\gamma_{l}$. Denote the two-dimensional unit vector
$\vec{e}=\vec{e}(\vec{x},\vec{y})$ as
\begin{equation}
e^{a}e^{a}=1\;\;\;(a=1,2;\vec{e}\perp\vec{m}).
\end{equation}
Then according to Ref. \cite{liuxin13}, $A_{\mu}$ can be decomposed in
terms of this two-dimensional vector $e^{a}$:
$A^i_{\mu}=\epsilon^{ab}e^{a}\partial_{\mu}e^{b}-\partial_{\mu}\varphi$;
$A^j_{\mu}=\epsilon^{ab}e^{a}\partial_{\mu}e^{b}-\partial_{\mu}\theta$,
where $\varphi$ and $\theta$ are phase factors. Since the terms
$\partial_{\mu}\varphi$ and $\partial_{\mu}\theta$ contribute nothing to
the integral $N(F^i,F^j,F^k)$, $A_{\mu}$ can in fact be expressed as
\begin{equation}\label{ee}
A^i_{\mu}=\epsilon^{ab}e^{a}\partial_{\mu}e^{b},\;\;\;
A^j_{\mu}=\epsilon^{ab}e^{a}\partial_{\mu}e^{b}.
\end{equation}
Substituting Eq. (\ref{ee}) into Eq. (\ref{linking3}) and considering
$\vec{x}$ and $\vec{y}$ are two points, respectively, on the knots
$\gamma_{k}$ and $\gamma_{l}$, we have
\begin{equation}\label{linking4}
N(F^i,F^j,F^k)=\sum^{N}_{k=1}\sum^{N}_{l=1}W_{k}W_{l}\oint_{\gamma_{k}}\oint_{\gamma_{l}}
\epsilon_{ab}\partial_{\mu}e^{a}(\vec{x},\vec{y})
\partial_{\nu}e^{b}(\vec{x},\vec{y})dx^{\mu}\wedge{dy^{\nu}}.
\end{equation}
Using the relation
$\epsilon^{ab}\partial_{\mu}e^{a}\partial_{\nu}e^{b}=\frac{1}{2}\vec{m}\cdot
(\partial_{\mu}\vec{m}\times\partial_{\nu}\vec{m})$, Eq. (\ref{linking4})
is just written as
\begin{equation}\label{linking5}
N(F^i,F^j,F^k)=\frac{1}{2}\sum^{N}_{k=1}\sum^N_{l=1}W_{k}W_{l}\oint_{\gamma_{k}}
\oint_{\gamma_{l}}\vec{m}\cdot(\partial_{\mu}\vec{m}
\times\partial_{\nu}\vec{m})dx^{\mu}\wedge{dy^{\nu}}.
\end{equation}
Substituting Eq. (\ref{m}) into Eq. (\ref{linking5}), one can prove that
\begin{eqnarray}
&\frac{1}{4\pi}\sum^{N}_{k=1}\oint_{\gamma_{k}}\oint_{\gamma_{l}}\vec{m}\cdot
(\partial_{\mu}\vec{m}\times\partial_{\nu}\vec{m})dx^{\mu}\wedge{dy^{\nu}}\\\nonumber
&=\frac{1}{4\pi}\epsilon_{\mu\nu\lambda}\oint_{\gamma_{k}}dx^{\mu}\oint_{\gamma_{l}}dy^{\nu}
\frac{(x^{\lambda}-y^{\lambda})}{||\vec{x}-\vec{y}||^{3}}=L(\gamma_{k},\gamma_l)(k\neq{l}),
\end{eqnarray}
where $L(\gamma_{k},\gamma_l)(k\neq{l})$ is the Gauss linking number
between $\gamma_{k}$ and $\gamma_l$ \cite{Gausslinking}. Therefore, we
arrive at the important result
\begin{equation}
I=2\pi\sum^{N}_{k=1}\sum^N_{l=1}W_{k}W_{l}L(\gamma_{k},\gamma_l).
\end{equation}
This precise expression just reveals the relationship between the new
invariant and the linking number of the knots family
\cite{knotfamily1,knotfamily2}. Since the linking numbers are invariant
characteristic numbers of the knotlike closed curves in topology,
$N(F^i,F^j,F^k)$ is an important invariant required to describe the linkage
of knotlike vortex lines in the $U(1)$ gauge field.

\section{conclusion}\label{section4}

In this paper, using the decomposition of $U(1)$ gauge potential theory and
the $\phi$-mapping topological current theory, we obtain the topological
structure of the knot invariant in the three-dimensional $U(1)$ gauge field
and that of the link invariant in four-dimensional case. We show that the
action $I(F^i,F^j)$ is a knot invariant for the knotlike vortex lines
inhering in the $U(1)$ gauge field. Furthermore, by generalizing to
four-dimensional case, we construct a new action $N(F^i,F^j,F^k)$ and
discuss the topological properties of the new action. It is pointed out
that the new action $N(F^i,F^j,F^k)$ is a topological invariant for the
knotlike vortex lines existing in the zeros of $\vec\phi$ field. Finally,
by introducing the Gauss mapping, we find out the relationship between the
new invariant $N(F^i,F^j,F^k)$ and the Gauss linking number. Therefore, we
reveal that the new action is a link invariant for knotlike vortex lines in
four dimensions.

\section{acknowledgment}

This work was supported by the National Natural Science Foundation
of China and the Doctoral Foundation of the People's Republic of
China.

\end{document}